\documentclass{aa}  

\usepackage{graphicx}
\usepackage{txfonts}
\usepackage{tabularx}

\newcommand\clearrow{\global\let\rowmac\relax}
\clearrow

\usepackage{lscape}

\begin{document}

   \title{Uncertainties in the $^{18}$F(p,\,$\alpha$)$^{15}$O reaction rate in classical novae}


   \author{D. Kahl
          \inst{1,}\inst{2}
          \and
	  J. Jos\'e
	  \inst{3,}\inst{4}
	  \and
	  P. J. Woods
	  \inst{1}
          }

   \institute{School of Physics \& Astronomy, James Clerk Maxwell Building, University of Edinburgh, Edinburgh EH9 3FD, UK
	\and Extreme Light Infrastructure -- Nuclear Physics, Horia Hulubei National Institute for R\&D in Physics and Nuclear Engineering (IFIN-HH), 077125 Bucharest-M\u{a}gurele, Romania \\
              \email{david.kahl@eli-np.ro}
         \and
              Departament de F\'isica, Universitat Polit\`ecnica de Catalunya, EEBE, 08019 Barcelona, Spain
         \and
	      Institut d’Estudis Espacials de Catalunya (IEEC), E-08034 Barcelona, Spain
             }

   \date{Last updated: June 4, 2021}

 
  \abstract
   {
     Direct observation of $\gamma$-ray emission from the decay of $^{18}$F ejected in classical nova outbursts remains a major focus of the nuclear astrophysics community.
     However, modeling the abundance of ejected $^{18}$F, and thus the predicted detectability distance of a $\gamma$-ray signal near 511~keV
     emitted from these transient thermonuclear episodes, 
     is hampered by significant uncertainties in our knowledge of the key $^{18}$F(p,\,$\alpha$) reaction rate.
   }
   {
     We analyze uncertainties in the most recent nuclear physics experimental results employed to calculate the $^{18}$F(p,\,$\alpha$) reaction rate.
     Our goal is to determine which uncertainties have the most profound influence on the predicted abundance of $^{18}$F ejected from novae, in order 
     to guide future experimental works.
   }
   {
     We calculated a wide range of $^{18}$F(p,\,$\alpha$) reaction rates using $R$-Matrix formalism, allowing us to take into account all interference effects.
     Using a selection of 16 evenly-spaced rates over the full range, we performed 16 new hydrodynamic nova simulations. 
   }
   {
     We performed one of the most thorough theoretical studies of the impact of the $^{18}$F(p,\,$\alpha$) reaction in classical novae to date.
     The $^{18}$F(p,\,$\alpha$) rate remains highly uncertain at nova temperatures, resulting in a factor $\sim\!10$ uncertainty in the predicted abundance of
     $^{18}$F ejected from nova explosions.
     We also found that the abundance of $^{18}$F may be strongly correlated with that of $^{19}$F.
   }
   {
     Despite numerous nuclear physics uncertainties affecting the $^{18}$F(p,\,$\alpha$) reaction rate, which are dominated by unknown interference signs between $1/2^{+}$ and $3/2^{+}$ resonances, future experimental work should focus on firmly and precisely 
     determining the directly measurable quantum properties of the subthreshold states in the compound nucleus $^{19}$Ne near 6.13 and 6.29~MeV.
   }


   \maketitle
%

\section{Introduction}

Classical novae are thermonuclear explosions that take place in the envelopes of accreting white dwarfs in stellar, short orbital period, binary systems (see \cite{StarrfieldIliadis2008,JoseShore2008,Jose2016} for reviews).
They have been systematically observed in all wavelengths, ranging from radio to $\gamma$-rays.
However, while novae have been predicted to emit $\gamma$-rays of specific energies around 1~MeV, associated to electron-positron annihilation (e.g., from $^{18}$F$(\beta^{+}\nu)$) and to nuclear decay (e.g., $^{22}$Na at 1.275~MeV and $^{26}$Al at 1.809~MeV), they have only been detected in $\gamma$-rays for energies exceeding 100~MeV.
In symbiotic binary systems, such as V407 Cyg, this high-energy emission has been attributed to shock acceleration in the ejected shells after interaction with the dense wind of the red giant companion.
The emission reported from several classical novae involving less evolved stellar companions (e.g., V1324 Sco, V959 Mon, V339 Del, \& V1369 Cen \citep{2014Sci...345..554A} has been attributed to internal shocks in the ejecta.
In this scenario, $\gamma$-rays are produced either through a hadronic process, in which accelerated protons collide with matter to create neutral pions that decay into $\gamma$-ray photons, or through a leptonic process, in which visible or infrared photons reach high energies in the interaction with high-energy electrons.
While the hadronic process seems to be favored, the exact nature of the mechanism responsible for $\gamma$-ray production remains unknown.

The most intense $\gamma$-ray signal predicted for classical novae corresponds to the 511-keV line due to electron-positron annihilation in the expanding ejecta, together with a lower energy continuum (powered both by positronium decay and by Comptonization of 511-keV photons) with a cut-off at $\sim$20--30 keV due to photoelectric absorption \citep{1974ApJ...187L.101C,1987ApJ...323..159L,1998MNRAS.296..913G,1999ApJ...526L..97H}.
As the expanding envelope begins to become transparent to $\gamma$ radiation after around $\sim\!2$~hrs, the main contributor to the bulk of positrons is $^{18}$F, which decays with a characteristic half-life, $t_{1/2}$, of about 110~min.
Therefore, all nuclear processes involved in the synthesis and destruction of $^{18}$F are of paramount importance to predict the expected amount of $^{18}$F that survives the explosion, and powers the predicted prompt 511-keV line and the lower-energy continuum.
Synthesis of $^{18}$F in novae is driven by the CNO-cycle reaction $^{16}$O(p,\,$\gamma$)$^{17}$F, which is either followed by $^{17}$F(p,\,$\gamma$)$^{18}$Ne($\beta^+$)$^{18}$F or by $^{17}$F($\beta^+$)$^{17}$O(p,\,$\gamma$)$^{18}$F.
Because of the relatively large half-life of $^{18}$F, its dominant destruction channel is mainly $^{18}$F(p,\,$\alpha$)$^{15}$O, plus a minor contribution from $^{18}$F(p,\,$\gamma$)$^{19}$Ne.
The most uncertain nuclear process involved in the creation and destruction of $^{18}$F for nova conditions is $^{18}$F(p,\,$\alpha$)$^{15}$O (\cite{2006NuPhA.777..550J,Jose2016}; and references therein).

We begin in Sect.~\ref{sec:nuclear} with a review and analysis of the present nuclear physics uncertainties affecting the $^{18}$F(p,\,$\alpha$) reaction, including the latest experimental results.
Using these evaluated experimental nuclear uncertainties, in Sect.~\ref{sec:rate} we calculate a range of reaction rates and vary these within new hydrodynamic nova simulations.
Subsequently, in Sect.~\ref{sec:results} we discuss and interpret all observed differences in our model outputs.
In Sect.~\ref{sec:conclusions}, we conclude by providing a roadmap to guide future nuclear experiments.

\section{Nuclear physics uncertainties}
\label{sec:nuclear}
The $^{18}$F(p,\,$\alpha$)$^{15}$O reaction plays a critical role in nova explosions, yet insufficient experimental information is available to calculate a reliable, precise rate for this reaction.
A quarter of a century ago, the first pioneering direct measurements were performed using $^{18}$F radioactive beams \citep{1995PhLB..353..184C,1995PhRvC..52..460R}.
However, the cross section at the lowest temperatures achieved in novae ($T_{8}\approx$ 1, where $T_{8}$ is in units of $10^{8}$~K) remains highly uncertain, because it is too small for direct measurement, even with presently available radioactive beam intensities.
Consequently, the structure of the $^{19}$Ne compound nucleus has also been extensively studied in order to indirectly estimate the $^{18}$F(p,\,$\alpha$) stellar reaction rate; for previous evaluations of the $^{18}$F(p,\,$\alpha$) rate, see, e.g., \cite{2007PhRvC..75e5809N}, \cite{2010NuPhA.841...31I}, \cite{2017ApJ...846...65L}, \cite{2019EPJA...55....4K}, \cite{2019PhRvL.122e2701H}, and \cite{2020PhRvC.102d5802H}.
As our understanding of excited states near the proton threshold in $^{19}$Ne continues to progress, it is necessary to address the latest findings and any discrepancies.  
Several new experimental works were published recently \citep{2019EPJA...55....4K,2019PhRvL.122e2701H,2019PhRvC..99c4301L}, and it is timely to review the status of nuclear physics uncertainties to examine which ones most strongly influence the $^{18}$F(p,\,$\alpha$) rate in classical novae.
Here we provide the first self-consistent evaluation of the $^{18}$F(p,\,$\alpha$) reaction rate combining all these latest results in a hydrodynamic nova model.

The $^{18}$F(p,\,$\alpha$) reaction rate at the highest nova temperatures ($T_{8}\approx 4$) is dominated by a $J^{\pi}=3/2^{-}$ (orbital angular momentum $\ell=1$) state at a resonance energy $E_{r}=332$~keV and a broad $3/2^{+}$ ($\ell=0$) resonance at 665~keV \citep{2002PhRvL..89z2501B,2001PhRvC..63f5802B}; we adopt the precisely measured properties of these states in the present work.
At lower nova temperatures ($T_{8}\lesssim 3$), $1/2^{+}$ and $3/2^{+}$ ($\ell=0$) states predicted near the proton threshold in $^{19}$Ne, including subthreshold resonances, can significantly influence the reaction rate by interfering with higher-lying, broad $1/2^{+}$ and $3/2^{+}$ states \citep{2009PhRvL.102p2503D,2006PhRvC..74a2801C}.
A broad $1/2^{+}$ state was reported, but not resolved, near $E_{r}\approx1.5$~MeV in several works \citep{2009PhRvL.102p2503D,2012PhRvC..85c7601A,2012PhRvC..85b2801M}.
Most recently, an $\ell=0$ state with a total width of $\Gamma=130(10)$~keV was firmly observed at 1.38(3)~MeV by \cite{2019EPJA...55....4K}.

   \begin{figure}
   \centering
   \includegraphics[width=4cm, clip=true, trim=0 0 0 0]{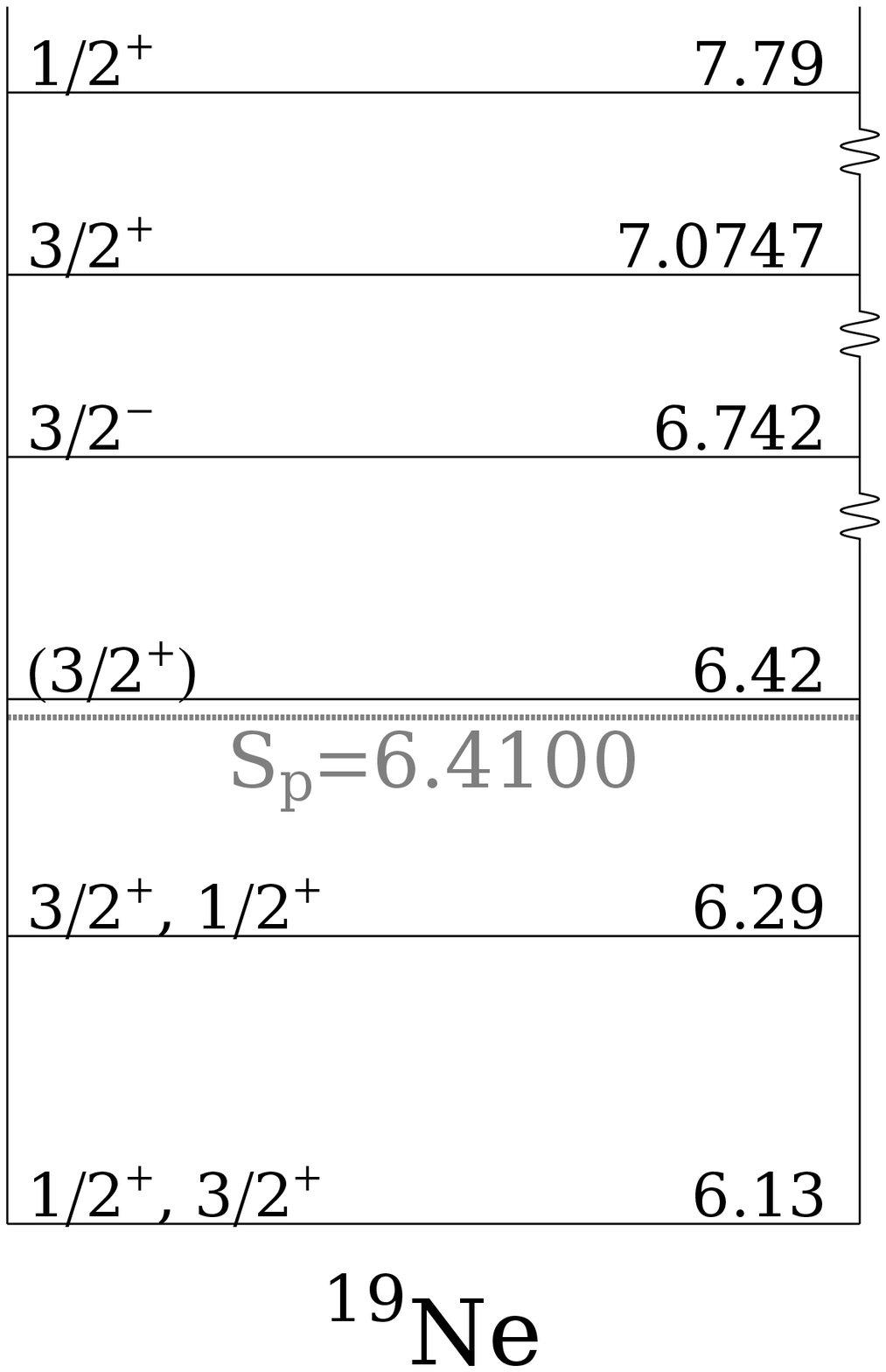}
      \caption{Partial level scheme of $^{19}$Ne, with emphasis on s-wave states affecting the $^{18}$F(p,\,$\alpha$) reaction rate.  See Table~\ref{tab:nuclear} for details.
              }
         \label{fig:levels}
   \end{figure}
\begin{table*}
 \caption{
  Resonances parameters and uncertainties used in the $^{18}$F(p,\,$\alpha)$ reaction rate calculations. 
}\label{tab:nuclear}
\centering
 \begin{tabular}{>{\rowmac}c>{\rowmac}c>{\rowmac}c>{\rowmac}c>{\rowmac}c<{\clearrow}}
      \hline\hline\noalign{\smallskip}
      $E_{\rm ex}$& $E_{\rm c.m.}$  & $J^{\pi}$ & ANC or $\Gamma_{\rm p}$     & $\Gamma_{\alpha}$  \\  
      (MeV)          & (keV)           &           & (fm$^{-1/2}$ or keV) & (keV) \\  
\noalign{\smallskip}\hline\noalign{\smallskip}
                  6.132(5)~~\tablefootmark{a}       & $-278(5)$       & $1/2^{+}$, $3/2^{+}$ &  $\leq$8~~\tablefootmark{b}, $\leq$6~~\tablefootmark{b}            & $8.6~~\tablefootmark{c}\sim16$~~\tablefootmark{d}\!, $0.74~~\tablefootmark{c}\sim8.5$~~\tablefootmark{e} \\  
                  $6.286(4)$~~$^{f,\,g}$      & $-124(4)$       & $3/2^{+}$, $1/2^{+}$ &  59~~\tablefootmark{h},  83.5$~~\tablefootmark{h}$             & $1.00~~\tablefootmark{c}\sim11.6$~~\tablefootmark{e}\!, $11.7~~\tablefootmark{c}\sim16$~~\tablefootmark{d} \\  
6.416~~\tablefootmark{b}, 6.423~~\tablefootmark{g}  & 6, 13            & (3/2$^{+}$) & $\leq$4.2$\times 10^{-45}$~~\tablefootmark{b,\,h}\!, $\leq$3.9$\times 10^{-29}$~~\tablefootmark{g,\,h} & $\leq$0.5~~\tablefootmark{c} \\
6.742~~$^{i}$          & 332             & 3/2$^{-}$ & 2.22$\times$10$^{-3}$~~\tablefootmark{i}   & 5.2~~\tablefootmark{c} \\
7.0747~~$^{j}$         & 664.7             & 3/2$^{+}$ & 15.2~~\tablefootmark{j}                   & 23.8~~\tablefootmark{j} \\
                  7.79(3)~~$^{b}$        & 1380(30)            & $1/2^{+}$ & 83$^{+56}_{-82}$~~\tablefootmark{b,\,k}                & 47$^{-46}_{+92}$~~\tablefootmark{b,\,k} \\  
\noalign{\smallskip}\hline
 \end{tabular}
\tablefoot{
  The asymptotic normalization coefficient (ANC) is used for subthreshold states ($E_{\rm c.m.}<0$) and the proton partial width, $\Gamma_{\rm p}$, otherwise.
  Values tabulated firmly were not varied in the present work.
  In the case of values separated by a comma, corresponding quantities are tabulated in the same order throughout a row.
  See the text and Fig.~\ref{fig:levels}. \\
\tablefoottext{a}{\cite{2013PhRvL.110c2502L}} 				
\tablefoottext{b}{\cite{2019EPJA...55....4K}}  				
\tablefoottext{c}{\cite{2005PhRvC..71a8801B,2007PhRvC..75e5809N}}  	
\tablefoottext{d}{\cite{2017PhRvC..96d4317T}}  				
\tablefoottext{e}{\cite{2019PhRvC..99c4301L}}  				
\tablefoottext{f}{\cite{2015PhRvC..92e5806P}}				
\tablefoottext{g}{\cite{2019PhRvL.122e2701H}} 				
\tablefoottext{h}{\cite{2011PhRvC..83e2801A,2011PhRvC..84e4611A}}, 	
\tablefoottext{i}{\cite{2002PhRvL..89z2501B}} 				
\tablefoottext{j}{\cite{2001PhRvC..63f5802B}} 				
\tablefoottext{k}{\cite{2012PhRvC..85c7601A}} 				
}
\end{table*}

To explore the magnitude of quantum interference effects on the $^{18}$F(p,\,$\alpha$) cross section, we carefully evaluate uncertainties in our knowledge of s-wave states in $^{19}$Ne close to the proton separation energy $S_{\rm p}=6.4100(5)$~MeV \citep{2017ChPhC..41c0003W}.
Because $^{19}$F (the mirror nucleus of $^{19}$Ne) is stable, its excitation structure has been comparatively well understood (see, e.g., \cite{1995NuPhA.595....1T}), including a comprehensive, high-resolution experimental and theoretical investigation of electron inelastic scattering \citep{1985PhRvC..32.1127B}.
Near the proton threshold energy in $^{19}$Ne, there is one $1/2^{+}$ state and two $3/2^{+}$ states known in $^{19}$F located at 6.255, 6.497, and 6.528~MeV, respectively.
We therefore focus our attention on the location and properties of three such s-wave states in $^{19}$Ne.
The values adopted in the present work are shown in Fig.~\ref{fig:levels} \& Table~\ref{tab:nuclear} and discussed presently.

Most recently, \cite{2019PhRvL.122e2701H,2020PhRvC.102d5802H} considered the astrophysical impact of uncertainties in the resonance properties on the $^{18}$F(p,\,$\alpha$) reaction rate by deriving attributes of multiple resonances in $^{19}$Ne from the same analog state in $^{19}$F.
Unfortunately, such an approach can lead to unphysical results. 
In the present work, all calculations are self-consistent, i.e., they are in agreement with the overall known structure and properties of $^{19}$F.
For example, we only pair the properties of the known 6.528~MeV state in $^{19}$F with just one candidate analog state in $^{19}$Ne for each calculation, whereas these properties are applied simultaneously to four resonances in $^{19}$Ne by \cite{2020PhRvC.102d5802H}.
This, for example, impacts on the potential importance of the recently observed $\ell=0$ subthreshold resonance at 6.13~MeV in $^{19}$Ne by \cite{2019EPJA...55....4K}.

Despite tabulated mirror pairings,\footnote{\cite{2020PhRvC.102d5802H} show a firm value for $\Gamma_{\alpha}$ of the $^{19}$F state at 6.497~MeV in their Table 1, where there is only an upper limit (see \cite{2007PhRvC..75e5809N} cited therein).  They also do not consider the recent value of $\Gamma_{\alpha}=13.9$~keV for the 6.528-MeV state in $^{19}$F \citep{2019PhRvC..99c4301L}.} \cite{2020PhRvC.102d5802H} clearly obtained the alpha partial width, $\Gamma_{\alpha}$, for both their suggested $(3/2^{+})$ states in $^{19}$Ne from the broader 6.528~MeV state in $^{19}$F. 
They further adopted a $(5/2^{-})$ resonance with $\Gamma_{\alpha}$ and the proton partial width, $\Gamma_{\rm p}$, derived from the same 6.528~MeV state in $^{19}$F (see Appendix~A \& \cite{2013PhRvL.110c2502L}).
\cite{2020PhRvC.102d5802H} then evaluated subsequent `inclusion' of the 6.13~MeV state in $^{19}$Ne (with its properties derived from the same 6.528-MeV state in $^{19}$F) has a less than $\pm 7.5\%$ influence on their reaction rate limits over nova temperatures. 
Those authors justified not `including' this $\ell=0$ state because ``no $\gamma$-ray transitions were observed from this state, and no significant strength to populate this state was observed in the $^{18}$F(d,\,n)$^{19}$Ne measurements \citep{2011PhRvC..83e2801A},'' yet they adopted other states which do not satisfy the same criteria.
We demonstrate in Fig.~\ref{fig:rate_nonL0} that the sum of four $\ell>0$ resonances included by \cite{2019PhRvL.122e2701H,2020PhRvC.102d5802H} exhibit a smaller influence than 7.5\% near peak nova temperatures, resulting in an inconsistent methodology.

In the present work, for the neighboring $3/2^{+}$ states at 6.497 and 6.528~MeV in $^{19}$F, one of which has a measured alpha strength and one just an upper limit from experiment, we only allow one of these values per candidate analog state in $^{19}$Ne.
For near- and sub-threshold resonances in the $^{18}$F(p,\,$\alpha$) reaction, $\Gamma$ is dominated by $\Gamma_{\alpha}$.
We find the values of $\Gamma_{\alpha}$ chosen for a given state strongly affect the magnitude of interference with higher-lying, broad resonances and thus can significantly influence the $^{18}$F(p,\,$\alpha$) reaction rate.
In contrast, \cite{2019PhRvL.122e2701H,2020PhRvC.102d5802H} find $\Gamma_{\alpha}$ to be negligible by application of the isolated, narrow-resonance condition for the reaction A($i$,$j$)B as 
\begin{equation}
\gamma\equiv\frac{\prod_{i}\Gamma_{i}}{\sum_{i}\Gamma_{i}}\approx\Gamma_{j} \iff \Gamma_{j} \ll \Gamma_{i\neq j}, 
\end{equation}
where $\gamma$ is the reduced width \citep{RR88}. 
However, the simplification of Eq.~1 is inapplicable to the case of interference with a broad resonance of the same $J^{\pi}$.

\cite{2019EPJA...55....4K} unambiguously identified an $\ell=0$ state at 6.130(5)~MeV in $^{19}$Ne using a model-independent approach, and based on the systematics of Coulomb energy differences strongly favored pairing this state with the $1/2^{+}$ 6.255-MeV state in $^{19}$F.
This conclusion is supported by the significant strength from a $1/2^{+}$ resonance in $^{19}$Ne observed near $\sim 6.1$~MeV \citep{lairdphd,2002PhRvC..66d8801L} with a magnitude comparable to population of the $1/2^{+}$ 6.225-MeV state in analogous studies of $^{19}$F \citep{1970NuPhA.142..137G,1970NuPhA.155..644S}.
\cite{2017PhRvC..96d4317T} observed the signature of a $1/2$ state of ambiguous parity in $^{19}$Ne at 6.197(8)(50) [statistical and systematic uncertainties] with a large $\Gamma_{\alpha}$ of 16~keV.
In the present work, we adopt the excitation energy of 6.132(5) MeV from the highest resolution studies to observe this low-spin state \citep{2013PhRvL.110c2502L,2015PhRvC..92e5806P}.
As a possible $3/2^{+}$ assignment for this state cannot be definitively ruled out, we compare and contrast the implications of the $J$ assignment of the 6.132-MeV state.
Table \ref{tab:nuclear} shows that the resulting uncertainty in $\Gamma_{\alpha}$ of the 6.132-MeV state spans considerably different ranges depending upon the assumed $J$.

A strong $\ell=0$ component was observed by \cite{2011PhRvC..83e2801A,2011PhRvC..84e4611A} at 6.289~MeV, but subsequent studies led to the conflicting conclusions that the state was not low spin \citep{2013PhRvL.110c2502L} and that the state is $1/2^{+}$ \citep{2015PhLB..751..311B}. 
There is now compelling evidence of a doublet in this region. 
\cite{2015PhRvC..92e5806P} observed a structure broader than their experimental resolution ($\Gamma\approx 16$~keV) and identified the possibility of a doublet with peaks at 6.282 and 6.294~MeV.
Subsequently, \cite{2019EPJA...55....4K} demonstrated that the structure at $6.288(5)$~MeV cannot be attributed to a single, pure $\ell=0$ resonance.
Finally, \cite{2019PhRvL.122e2701H} observed a state at 6291.6(9)~keV they suggested to be ($11/2^{+}$).
Although such a high spin, subthreshold state would not contribute to the astrophysical reaction rate, its presence near a subthreshold $\ell=0$ state continues to pose an experimental challenge to simultaneously and unambiguously resolve.

The mirror partner of the remaining $\ell=0$ state in $^{19}$F is likely located above the proton threshold in $^{19}$Ne, but the situation is complicated as half a dozen states may exist within just 50~keV (see, e.g., \cite{2007PhRvC..75e5809N,2013PhRvL.110c2502L}).
\cite{2019PhRvL.122e2701H} tentatively suggest $(3/2^{+})$ assignments for two states above the proton threshold: a newly-proposed state at 6.423(3)~MeV and a previously-identified state at 6.441(3)~MeV.
Clearly three $\ell=0$ states in $^{19}$F cannot be paired with four states in $^{19}$Ne, and one of these tentative $(3/2^{+})$ pairings should be rejected.
\cite{2019EPJA...55....4K} observed a peak at 6.421(10) MeV which could include an $\ell=0$ component near the centroid, but an $\ell=0$ state at higher excitations is incompatible with that study and \cite{2013PhRvL.110c2502L}.
The $3/2^{+}$ state at 6.528~MeV in $^{19}$F has the larger $\Gamma_{\alpha}$, with values 4, 1.2, and 13.9~keV reported \citep{1961PhRv..122..232S,2005PhRvC..71a8801B,2019PhRvC..99c4301L}; its mirror partner in $^{19}$Ne is thus unlikely to be observed in the low-statistics $\gamma$-ray spectroscopy experiment of \cite{2019PhRvL.122e2701H}.
We therefore assume the mirror partner of the $3/2^{+}$ state at 6.497 MeV in $^{19}$F may be near 6.42~MeV in $^{19}$Ne, which is supported by perfect consistency in the observed $\gamma$ transitions \citep{1995NuPhA.595....1T,2019PhRvL.122e2701H}.

As for the possible contribution of $\ell>0$ resonances near the proton threshold, none have firm values for all their properties---besides many $J^{\pi}$ ambiguities, at least one of the partial widths for each resonance is an experimental upper limit (see, e.g., \cite{2007PhRvC..75e5809N,2013PhRvL.110c2502L}).
Nonetheless, we consider the possible impact of such $\ell>0$ states on the $^{18}$F(p,\,$\alpha$) reaction rate in novae (see Table~\ref{tab:sfactor2}). 

\section{Reaction rate and stellar model}
\label{sec:rate}
We calculated a wide range of $^{18}$F(p,\,$\alpha$) stellar reaction rates with the $R$-Matrix code \texttt{AZURE2} \citep{2010PhRvC..81d5805A,2014NIMPA.767..359M} based on the nuclear physics uncertainties presented in Table \ref{tab:nuclear}.
Our goal is to assess which specific nuclear quantities have the largest impact on nova observables in our model to guide future experimental measurements.
Based on our evaluation in Sect.~\ref{sec:nuclear}, we focused on, but did not limit ourselves to, variation of the following parameters within their uncertainties:
\begin{itemize}
\item Interference sign between the two $1/2^{+}$ states
\item Interference signs between up to three $3/2^{+}$ states
\item $J$ of the $\ell=0$ state at 6.13~MeV
\item Unknown proton ANC of the 6.13~MeV state
\item $J$ and $E_{\rm ex}$ of the $\ell=0$ state near 6.29~MeV
\item $\Gamma_{\alpha}$ for the mirror of the $3/2^{+}$, 6.528-MeV state in $^{19}$F
\item $\Gamma_{\alpha}$ of the $1/2^{+}$ subthreshold state
\item $E_{\rm ex}$ of a narrow $(3/2^{+})$ state just above $S_{\rm p}$
\item Influence of $\ell>0$ states near $S_{\rm p}$
\item $\Gamma$ as well as the $\Gamma_{\rm p}/\Gamma_{\alpha}$ ratio of the $1/2^{+}$ state near 7.8~MeV
\end{itemize}
We found variation in other properties to be implausible or inconsequential. 
Further details of the $R$-Matrix calculations are provided in Appendix~A.

\begin{figure}
   \centering
   \includegraphics[width=9.5cm, clip=true, trim=0 0 0 30]{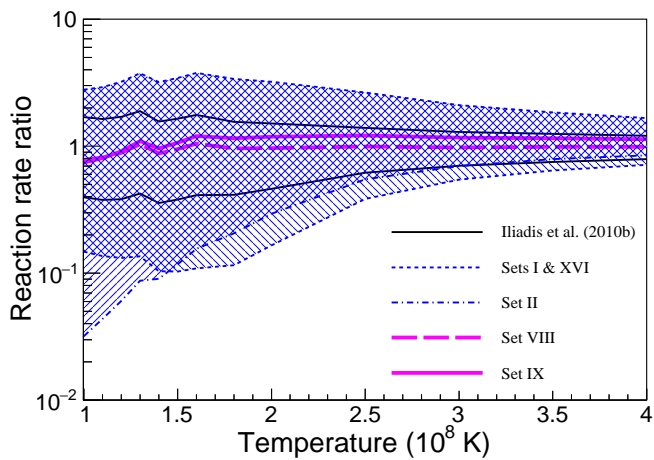}
      \caption{Range of reaction rates explored in the present work (see Tables~\ref{tab:nuclear} \& \ref{tab:rates}) normalized to the median rate of \cite{2010NuPhA.841...31I}; the low and high rates of \cite{2010NuPhA.841...31I} are also shown.  The recommended (median) rates from the present work are shown in magenta for Sets VIII \& IX, where the subthreshold spin ordering is $3/2^+,1/2^+$ or $1/2^+,3/2^+$, respectively.  The highest rate (XVI) represents a case of maximally constructive interference, while the lowest rates (I \& II) are obtained by maximally destructive interference.
              }
         \label{fig:ratenorm}
   \end{figure}

Considering this large set of codependent parameters, a Monte-Carlo approach, such as employed in \texttt{RatesMC} \citep{2010NuPhA.841....1L,2010NuPhA.841...31I,2010NuPhA.841..251I} is quite attractive.
However, the case of $^{18}$F(p,\,$\alpha$) presents a particular challenge owing to the possible influence of the interference between up to {\em three} resonances of the same $J^{\pi}$, namely the $3/2^{+}$ states.
For this reason, the previous evaluation of \cite{2010NuPhA.841...31I} considered only the interference between a pair of $3/2^{+}$ states, and \texttt{RatesMC} uses the simple analytic formula of Rolfs (Longland, priv. comm.).
Here we aim to improve upon the previous works by fully considering the latest experimental results using an $R$-matrix approach to handle multiple interference effects over a large and internally-consistent parameter space.
A comparison of the range of the present rates against those of \cite{2010NuPhA.841...31I} is shown in Fig.~\ref{fig:ratenorm}; the corresponding rates from the present work are listed in Table~\ref{tab:rates}.
We recommend the median rates (Sets XIII \& IX) obtained in the present work (see Fig.~\ref{fig:sfactor}).
After careful evaluation of the impact of each noted uncertainty on the stellar reaction rate, we selected a representative sample of 16 possible rates over the full range for a further analysis.

The astrophysical impact of these new $^{18}$F(p,\,$\alpha$) rates has been assessed through a series of 16 new hydrodynamic models of nova explosions. 
The simulations have been performed with the hydrodynamic, Lagrangian, finite-difference, time-implicit code {\tt SHIVA} \citep{1998ApJ...494..680J,Jose2016}. 
{\tt SHIVA} relies on the standard set of differential equations of stellar evolution and has been extensively used in the characterization of nova outbursts, Type-I X-ray bursts and sub-Chandrasekhar supernova explosions, for more than 20 years. 
{\tt SHIVA} uses a general equation of state that includes contributions from the degenerate electron gas, the ion plasma, and radiation. 
Coulomb corrections to the electron pressure are taken into account, and both radiative and conductive opacities are considered in the energy transport. 
Nuclear energy generation is obtained by means of a reaction network that contains 120  species (from $^1$H to $^{48}$Ti), linked through 630 nuclear processes, with updated rates from the STARLIB database (\cite{2013ApJS..207...18S} and Iliadis, priv. comm.). 
The accreted matter from the stellar companion, at a typical rate of $2 \times 10^{-10}$ M$_\odot$ yr$^{-1}$, is assumed to mix with material from the outer layers of the underlying white dwarf to a characteristic level of 50\%. 
In all the hydrodynamic simulations performed in this work, the mass of the (ONe) white dwarf hosting the explosion has been assumed to be 1.25 solar masses (and its initial luminosity $10^{-2}$ solar luminosities), and are identical, differing only in the specific prescription adopted for the $^{18}$F(p,\,$\alpha$) rate.

\section{Results}

\begin{table}
 \caption{
  Mean, mass-averaged abundances of $^{18,19}$F in the nova ejecta. 
}\label{tab:astro}
 \begin{tabular}{cccccc}
      \hline\hline\noalign{\smallskip}
Set & \multicolumn{2}{c}{Abundance}	& \multicolumn{3}{c}{Rate ($T_{8}$)} \\
&$^{18}$F~~\tablefootmark{a}	& $^{19}$F 	& 1 	& 2.5 	& 4  \\ \hline\noalign{\smallskip}
I&1.1$\times10^{-5}$	& 1.8$\times10^{-7}$	& 1.22$\times10^{-5}$      & $0.466$    & $96.5$ \\
II&9.3$\times10^{-6}$	& 1.2$\times10^{-7}$	& 2.63$\times10^{-6}$      & $0.664$    & $115$ \\
III&8.6$\times10^{-6}$	& 1.1$\times10^{-7}$	& 3.69$\times10^{-6}$      & $0.685$    & $116$ \\
IV&7.9$\times10^{-6}$	& 1.4$\times10^{-7}$	& 2.31$\times10^{-5}$      & $0.493$    & $95.7$ \\
V&7.8$\times10^{-6}$	& 9.9$\times10^{-8}$	& 6.57$\times10^{-6}$      & $0.702$    & $116$ \\
VI&7.0$\times10^{-6}$	& 8.9$\times10^{-8}$	& 8.83$\times10^{-6}$      & $0.740$    & $118$ \\
VII&2.9$\times10^{-6}$	& 4.0$\times10^{-8}$	& 6.05$\times10^{-5}$      & $1.16$     & $132$ \\
VIII&2.8$\times10^{-6}$	& 3.8$\times10^{-8}$	& 6.46$\times10^{-5}$      & $1.21$     & $134$ \\
IX&2.6$\times10^{-6}$	& 3.1$\times10^{-8}$	& 6.15$\times10^{-5}$      & $1.48$     & $154$ \\
X&2.4$\times10^{-6}$	& 2.9$\times10^{-8}$	& 7.03$\times10^{-5}$      & $1.55$     & $157$ \\
XI&1.9$\times10^{-6}$	& 2.3$\times10^{-8}$	& 1.10$\times10^{-4}$      & $1.82$     & $162$ \\
XII&1.8$\times10^{-6}$	& 2.2$\times10^{-8}$	& 1.17$\times10^{-4}$      & $1.89$     & $165$ \\
XIII&1.6$\times10^{-6}$	& 2.0$\times10^{-8}$	& 1.36$\times10^{-4}$      & $2.05$     & $170$ \\
XIV&1.4$\times10^{-6}$	& 1.6$\times10^{-8}$	& 1.58$\times10^{-4}$      & $2.51$     & $198$ \\
XV&1.3$\times10^{-6}$	& 1.6$\times10^{-8}$	& 1.77$\times10^{-4}$      & $2.53$     & $193$ \\
XVI&1.1$\times10^{-6}$	& 1.2$\times10^{-8}$	& 2.30$\times10^{-4}$      & $3.20$     & $225$ \\

\noalign{\smallskip}\hline
 \end{tabular}
\tablefoot{
  Calculated for each set of $^{19}$Ne resonance parameters considered for the $^{18}$F(p,\,$\alpha$) reaction rate, $N_{A}\langle\sigma v\rangle$, in cm$^{3}$mol$^{-1}$s$^{-1}$.
  Samples of the rates are provided at the representative temperatures of $T_{8}$=1, 2.5, and 4. 
  The sets are enumerated according to the $^{18}$F abundance for convenience. \\
\tablefoottext{a}{1 hour after $T_{\rm peak}$.}  
}
\end{table}

\label{sec:results}
   \begin{figure*}
   \centering
   \includegraphics[width=16cm, clip=true, trim=0 115 0 0]{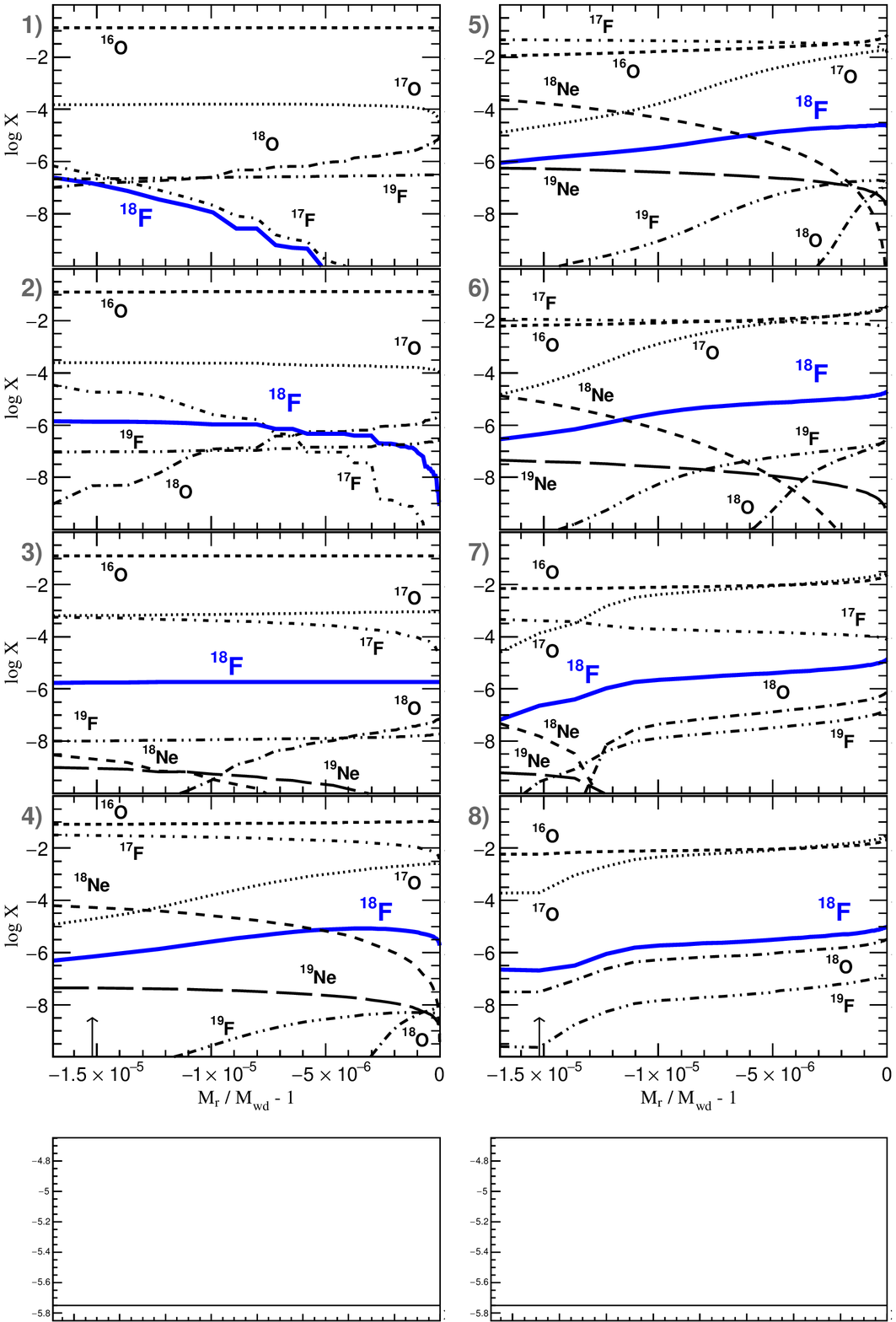}
   \caption{
     Mass fractions of selected isotopes within the expanding envelope at various stages of the explosion using our newly recommended $^{18}$F(p,\,$\alpha$) reaction rate Set IX.
The horizontal axes correspond to a relative mass
     coordinate where the origin is set at the surface of the envelope and where the arrows show the limit of the ejected material. The panels are
     labeled from top to bottom and from left to right and correspond to temperatures at the base of the accreted envelope of approximately $5\times10^{7}$ (Panel~1), $7\times10^{7}$ (Panel~2), $1\times10^{8}$ (Panel~3), $2\times10^{8}$ (Panel~4),
     $T_{\rm max}=2.4\times10^{8}$~K (Panel~5), respectively, and to the last phases of the evolution, when the nova envelope has already expanded to a size of 
     $R_{\rm wd}\sim 10^{9}$ (Panel~6), $10^{10}$ (Panel~7) and $10^{12}$~cm (Panel~8), respectively.
     We plot our results in a manner almost identical to Fig.~2 in \cite{2000A&A...357..561C} to facilitate comparison.
   }
   \label{fig:mass}%
    \end{figure*}

The main difference in our output nova model parameters is centered, as expected, around $^{18}$F.
As the energetics of the explosion are not affected by variations in the $^{18}$F(p,\,$\alpha$) reaction rate within current uncertainties, the results for other model quantities are identical, such as the total ejected mass, the peak temperature, $T_{\rm peak}$, achieved, etc.
The values shown in Table~\ref{tab:astro} correspond to mass-averaged (mean value) abundances of $^{18,19}$F in mass fractions, $X$, in the ejecta and representative temperature samples of the employed reaction rate.
The ejected abundances of all other isotopes examined varied between the models by at most a relative shift of $\pm 1$\%, which we adopt as the precision of sensitivity achieved in the present work.
While classical novae are thought to be a major source of $^{15}$N, $^{17}$O, and to some extent $^{13}$C (\cite{2006NuPhA.777..550J,Jose2016}; and references therein), we observed no significant variation in the abundances of any of these isotopes in the present study.
For example, the contribution of radiogenic $^{15}$N from the $^{18}$F(p,\,$\alpha$)$^{15}$O reaction is negligible, as our model consistently ejected $^{15}$N with an abundance of $3.39\times10^{-2}$, about four orders of magnitude larger than our predicted $\pm 5\times10^{-6}$ abundance shift in $^{18}$F.
It can be seen in Table~\ref{tab:astro} that the ejected abundances of $^{18}$F (1 hour after $T_{\rm peak}$) and $^{19}$F covary by about an order of magnitude and are inversely correlated with the respective $^{18}$F(p,\,$\alpha$) reaction rate.
As the observational detectability distance goes as the square root of the ejected mass fraction, our results indicate a factor of $\sim\!3$ uncertainty in the detectability distance of $\gamma$-rays from $^{18}$F decay in classical novae from the $^{18}$F(p\,$\alpha$) reaction rate alone.

The strong correlation of $^{18}$F and $^{19}$F abundances over a wide range of $^{18}$F(p,\,$\alpha$) reactions rates is consistent with its predominance over the $^{18}$F(p,\,$\gamma$)$^{19}$Ne reaction in novae.
$^{19}$Ne is very short-lived ($t_{1/2} \approx 18$~s) and decays to the only stable isotope of fluorine, $^{19}$F, which was indeed identified in the spectra of Nova Mon 2012 \citep{2013A&A...553A.123S}.
The abundance shifts in $^{19}$F can be understood, as the $^{18}$F(p,\,$\alpha$) rate limits the available $^{18}$F by flushing away material faster than the $^{18}$F(p,\,$\gamma$)$^{19}$Ne($\beta^{+}$)$^{19}$F sequence can occur. 
We emphasize that the true $^{18}$F(p,\,$\alpha$) rate remains quite uncertain.

Snapshots of the evolution of a number of representative isotopes of the O-Ne group, relevant in the synthesis and destruction of $^{18}$F (i.e., $^{16,17,18}$O, $^{17,18,19}$F, and $^{18,19}$Ne) are depicted in Fig.~\ref{fig:mass}. 
The initial mass fractions of these species throughout the envelope, at the onset
of the accretion stage, are as follows: $X(^{16}{\rm O})=0.13$, 
$X(^{17}{\rm O})=2.1 \times 10^{-6}$, $X(^{18}{\rm O})=1.2 \times 10^{-5}$, 
and $X(^{19}{\rm F})=3.1 \times 10^{-7}$.
As matter transferred from the secondary star piles up on the 
surface of the white dwarf star, compressional heating and nuclear
reactions increase the temperature (and pressure) of the accreted
envelope.  When the temperature at the base of the envelope achieves
$T_{\rm base} \sim 5 \times 10^7$ K (Fig.~\ref{fig:mass}, Panel~1), the main nuclear activity in the O-Ne mass region is dominated by the chain 
$^{16}$O(p,\,$\gamma$)$^{17}$F($\beta^+$)$^{17}$O(p,\,$\alpha$)$^{14}$N.
The main effect, at this stage, is the increase of  $^{17}$O, whose abundance
rises by two orders of magnitude at the base of the envelope, with respect
to its initial value. However, since the temperature achieved is not high enough to 
fuse a significant fraction of $^{16}$O, changes on most of the species are moderate. 
This mostly affects the build up of $^{17,18}$F, whose mass fractions reach
$\sim 10^{-7}$. Proton capture reactions onto $^{17,18}$F result also in a very modest
increase of both $^{18,19}$Ne ($<10^{-10}$, by mass). Convective transport is scarcely present at this stage, which explains the inhomogeneous chemical abundance pattern of the envelope (see, for instance, the larger abundance of $^{18}$F 
near the core-envelope interface in Panel~1). 

A similar behavior is observed when $T_{\rm base} \sim 7 \times 10^7$ K (Fig.~\ref{fig:mass},
Panel~2), with the same chain of reactions described in Panel~1 still 
dominating the nuclear activity in the envelope. Convection has already extended 
throughout larger areas of the envelope, which translates into a smoother chemical 
profiles across the envelope, for species such as $^{18}$F, which is synthesized close to the base of the envelope and efficiently carried away to the outer, cooler layers of the envelope. At this stage, $^{18}$F $\sim 10^{-6}$ ($^{17}$F $\sim 10^{-5}$) at the base of the envelope. Aside from $^{16}$O, the second most abundant isotope of the O-Ne group at this stage is still $^{17}$O.

When $T_{\rm base}$ reaches $10^8$ K (Fig.~\ref{fig:mass}, Panel~3), most of the accreted envelope becomes convective. This tends to homogenize the envelope, as shown by the nearly flat profiles of most species, particularly $^{16,17}$O, and $^{17,18,19}$F. The abundance of $^{16}$O is being reduced by (p,\,$\gamma$) reactions. $^{17}$O shows an increase, with a mass fraction ranging from $6.3 \times 10^{-4}$ (inner envelope) to $8.7 \times 10^{-4}$ (outer envelope). Here, the chain $^{16}$O(p,\,$\gamma$)$^{17}$F($\beta^+$)$^{17}$O dominates the destruction channel $^{17}$O(p,\,$\alpha$)$^{14}$N.
Both $^{17,18}$F have notoriously increased with respect to the previous panel, through $^{16,17}$O(p,\,$\gamma$)$^{17,18}$F reactions, while $^{19}$F has been severely depleted, mostly through $^{19}$F(p,\,$\alpha$)$^{16}$O.
$^{18,19}$Ne continue to rise in the innermost layers of the envelope by means of proton captures onto $^{17,18}$F.

When $T_{\rm base}$ reaches $2 \times 10^8$ K (Fig.~\ref{fig:mass}, Panel~4), $^{16}$O experiences a significant decrease by proton-capture reactions, which reduce its abundance by nearly a factor of two in the innermost layers of the envelope. This, in turn, induces a reduction in $^{17}$O (with $^{17}$O(p,\,$\alpha$)$^{14}$N reactions dominating over $^{17}$F($\beta^+$)$^{17}$O). With regard to $^{18}$F, its abundance is mostly dominated by the destruction channel  $^{18}$F(p,\,$\alpha$)$^{15}$O
at the hotter, inner regions of the envelope, while $^{17}$O(p,\,$\gamma$)$^{18}$F
and $^{18}$Ne($\beta^+$)$^{18}$F contribute to rise its mass fraction in the outer, cooler regions. As a result, $^{18}$F begins to show a steeper profile which, except for the outermost layers of the envelope, increases outward. At this stage, $^{18}$O and $^{19}$F have been severely reduced by proton captures, while a remarkable increase in $^{18,19}$Ne by several orders of magnitude, mainly driven by proton-capture reactions onto $^{17,18}$F, which at these temperature become faster than the corresponding $\beta^+$-decays. 

Shortly thereafter, the base of the envelope achieves a maximum temperature of $2.38 \times 10^8$ K (Fig.~\ref{fig:mass}, Panel~5). At this stage, the entire envelope has become fully convective. 
The $^{16}$O abundance continues to decrease by proton captures, and for the first time, a different species, $^{17}$F, becomes the most abundant O-Ne group element in most of the envelope. 
The nuclear path is still dominated by the $^{16}$O(p,\,$\gamma$)$^{17}$F($\beta^+$)$^{17}$O(p,\,$\alpha$)$^{14}$N chain. At this stage, the abundances of the different species reflect a competition between different destruction and creation modes, mostly governed by (p,\,$\gamma$) and (p,\,$\alpha$) reactions in the inner regions of the envelope, and by $\beta^+$-decays in the outer regions. 
With the exception of $^{16}$O, the other O-Ne group isotopes shown in Panel~5 experience a net increase.

A few minutes after the envelope attains peak temperature, the sudden release of energy from the very abundant, short-lived species $^{13}$N, $^{14,15}$O, and $^{17}$F, forces the envelope to expand. Hereafter, proton-capture reactions become confined to the innermost regions of the envelope, following a dramatic temperature decrease in the outer layers (Fig.~\ref{fig:mass}, Panels~6 \& 7). Simultaneously, convection begins to recede from the outer envelope shells. 

The final stages of the outburst (Fig.~\ref{fig:mass}, Panel~8), when a significant fraction
of the envelope has already achieved escape velocity, are dominated by the
release of nuclear energy by the $\beta^+$-decay processes $^{17,18}$F($\beta^+$)$^{17,18}$O. The mean $^{18}$F abundance in the ejecta for this Model, one hour after peak temperature, turns out to be $X(^{18}{\rm F})=2.6 \times 10^{-6}$. 
The most abundant isotopes of the O-Ne group in the ejecta are, however, $X(^{16}{\rm O})= 9.2 \times 10^{-3}$ and $X(^{17}{\rm O})= 7.9 \times 10^{-3}$. In addition, some amounts of $^{18}$O ($1.1 \times 10^{-6}$, by mass) and $^{19}$F ($3.1 \times 10^{-8}$, by mass), have been found for this Model, assuming that all radioactive isotopes have already decayed into the corresponding daughter nuclei.

We confirmed that the largest variation among the $^{18}$F(p,\,$\alpha$) reaction rates, and thus the ejected $^{18}$F, arises from the unknown interference signs between the $\ell=0$ resonances.
As for directly measurable quantum properties, we found that $\Gamma_{\alpha}$ of a broad $3/2^{+}$ subthreshold state has the most profound influence on both the maximum and minimum reaction rates. 
In the case of maximally destructive interferences, the lowest rates are always obtained when the 6.13~MeV state is assumed to be $1/2^{+}$. 
Conversely, the maximum rates are similar for either spin ordering for the $\ell=0$ subthreshold states, depending on the other assumptions. 
Our minimum rate is sensitive to the proton asymptotic normalization coefficient (ANC) of the 6.13~MeV state when it is $1/2^{+}$; here, a lower (but non-zero) value of 4~fm$^{-1/2}$ compared with the presumed experimental upper limit of 8~fm$^{-1/2}$ surprisingly leads to the largest effects, by shifting the dip of the destructive interference into the center of the astrophysical energy region. 
Although the broad $1/2^{+}$ state near 7.8~MeV is found to strongly affect the reaction rate by interference, we found that the rate is not sensitive to the adopted $\Gamma_{\rm p}/\Gamma_{\alpha}$ ratio nor the uncertainty in $E_{\rm ex}$ for this state.
Finally, a $3/2^{+}$ state above the proton threshold can have a modest impact on the rate, depending on its exact energy and if $\Gamma_{\alpha}$, $\Gamma_{\rm p}$, and $E_{\rm ex}$ are simultaneously taken to be near the maximal values consistent with experimental measurements, although we consider such a scenario to be unlikely.

We found that the possible impact of $\ell>0$ states (e.g., the ones assumed in \cite{2019PhRvL.122e2701H}) near $S_{\rm p}$ to the maximum $^{18}$F(p,\,$\alpha$) rate is in general trivial in relation to other nuclear uncertainties, the magnitude of their effect on the reaction rate being comparable to the least-influential uncertainties of $\ell=0$ resonances we considered (see Appendix~A).
For the minimum rate, setting a given partial width with only an upper limit to an arbitrarily small value is compatible with experimental results. 

\section{Conclusions}
\label{sec:conclusions}

We performed a thorough analysis of the impact of the presumed $^{18}$F(p,\,$\alpha$) reaction rate on the predicted abundance of the astronomical observable $^{18}$F produced in classical novae.
The reaction rates were obtained using the $R$-Matrix method, where the uncertainties in the nuclear input parameters were derived from a self-consistent evaluation of all available nuclear physics data, including the latest experimental results.
An advantage of our approach was the ability to include all quantum interference effects on the reaction rate.
The range of reaction rates we explored turned out to be quite large, and we selected 16 evenly-spaced rates over our full range for further analysis.
Using these new rates, we performed 16 (otherwise identical) hydrodynamic nova simulations, and we presented and analyzed the only differences in model results we observed.
The uncertainty in the observational distance arising from the decay of ejected $^{18}$F from novae is about a factor of three.
We conclude by emphasizing the extant nuclear physics quantities which need to be known more precisely to significantly reduce this astronomical uncertainty: 
\begin{itemize}
\item Our lack of precise knowledge of the properties of the $\ell=0$ states near 6.13 \& 6.29 dominate the uncertainties.
\item The proton ANC of the 6.13~MeV state.
\item More definitive spin assignments on the $\ell=0$ states near 6.13 \& 6.29~MeV, ideally with $<5$~keV resolution.
\item A stronger upper-limit on $\Gamma_{\alpha}$, and thus $\Gamma$, of the putative $3/2^{+}$ state near 6.42~MeV.
\end{itemize}
In the future, it will be interesting to investigate the uncertainties in other nuclear reactions affecting $^{19}$F abundances in novae to determine if it can be used as an observational tracer for the $^{18}$F(p,\,$\alpha$) reaction.

\begin{acknowledgements}
D.K. and P.J.W. are appreciative of funding from the UK STFC.
J.J. acknowledges partial support through the Spanish MINECO grant AYA2017-86274-P, E.U. FEDER funds, and through the AGAUR/Generalitat de Catalunya grant SGR-661/2017. 
This article benefited also from discussions within the “ChETEC” COST Action (CA16117).
\end{acknowledgements}

%
%
\bibliographystyle{aa}
\bibliography{library}

\longtab{
\renewcommand\thetable{B.1} 
\begin{longtable}{cccccc}
\caption{\label{tab:rates} Thermonuclear reaction rates for the $^{18}$F(p,\,$\alpha$) sets shown in Fig.~\ref{fig:ratenorm}.  }\\
\hline\hline
\hline\hline
$T_{8}$ & Set I & Set II & Set VIII & Set IX \tablefootmark{$\dagger$} & Set XVI \\
   & \multicolumn{2}{c}{Minima} & \multicolumn{2}{c}{Medians} & Maximum \\
\hline\noalign{\smallskip}
\endfirsthead
\caption{continued.}\\
\hline\hline
$T_{8}$ & Set I & Set II & Set VIII & Set IX \tablefootmark{$\dagger$} & Set XVI \\
   & \multicolumn{2}{c}{Minima} & \multicolumn{2}{c}{Medians} & Maximum \\
\hline\noalign{\smallskip}
\endhead
\hline
\endfoot
0.010 & 7.336$\times 10^{-65}$ & 9.379$\times 10^{-65}$ & 1.228$\times 10^{-64}$ & 5.253$\times 10^{-65}$ & 4.664$\times 10^{-65}$\\
0.020 & 6.493$\times 10^{-49}$ & 3.201$\times 10^{-48}$ & 1.108$\times 10^{-48}$ & 4.881$\times 10^{-49}$ & 4.984$\times 10^{-49}$\\
0.030 & 3.459$\times 10^{-41}$ & 3.938$\times 10^{-39}$ & 6.006$\times 10^{-41}$ & 2.710$\times 10^{-41}$ & 3.522$\times 10^{-39}$\\
0.040 & 2.546$\times 10^{-36}$ & 6.399$\times 10^{-34}$ & 4.490$\times 10^{-36}$ & 2.069$\times 10^{-36}$ & 6.504$\times 10^{-34}$\\
0.050 & 7.290$\times 10^{-33}$ & 9.337$\times 10^{-31}$ & 1.305$\times 10^{-32}$ & 6.129$\times 10^{-33}$ & 1.048$\times 10^{-30}$\\
0.060 & 3.132$\times 10^{-30}$ & 1.295$\times 10^{-28}$ & 5.683$\times 10^{-30}$ & 2.718$\times 10^{-30}$ & 1.758$\times 10^{-28}$\\
0.070 & 3.940$\times 10^{-28}$ & 5.091$\times 10^{-27}$ & 7.245$\times 10^{-28}$ & 3.524$\times 10^{-28}$ & 9.911$\times 10^{-27}$\\
0.080 & 2.115$\times 10^{-26}$ & 1.025$\times 10^{-25}$ & 3.939$\times 10^{-26}$ & 1.947$\times 10^{-26}$ & 3.253$\times 10^{-25}$\\
0.090 & 6.098$\times 10^{-25}$ & 1.445$\times 10^{-24}$ & 1.150$\times 10^{-24}$ & 5.773$\times 10^{-25}$ & 7.250$\times 10^{-24}$\\
0.100 & 1.099$\times 10^{-23}$ & 1.589$\times 10^{-23}$ & 2.098$\times 10^{-23}$ & 1.068$\times 10^{-23}$ & 1.136$\times 10^{-22}$\\
0.110 & 1.371$\times 10^{-22}$ & 1.395$\times 10^{-22}$ & 2.651$\times 10^{-22}$ & 1.369$\times 10^{-22}$ & 1.304$\times 10^{-21}$\\
0.120 & 1.277$\times 10^{-21}$ & 9.915$\times 10^{-22}$ & 2.498$\times 10^{-21}$ & 1.307$\times 10^{-21}$ & 1.149$\times 10^{-20}$\\
0.130 & 9.359$\times 10^{-21}$ & 5.825$\times 10^{-21}$ & 1.853$\times 10^{-20}$ & 9.821$\times 10^{-21}$ & 8.107$\times 10^{-20}$\\
0.140 & 5.630$\times 10^{-20}$ & 2.897$\times 10^{-20}$ & 1.128$\times 10^{-19}$ & 6.052$\times 10^{-20}$ & 4.744$\times 10^{-19}$\\
0.150 & 2.868$\times 10^{-19}$ & 1.248$\times 10^{-19}$ & 5.810$\times 10^{-19}$ & 3.156$\times 10^{-19}$ & 2.369$\times 10^{-18}$\\
0.160 & 1.268$\times 10^{-18}$ & 4.741$\times 10^{-19}$ & 2.599$\times 10^{-18}$ & 1.429$\times 10^{-18}$ & 1.032$\times 10^{-17}$\\
0.180 & 1.751$\times 10^{-17}$ & 5.007$\times 10^{-18}$ & 3.669$\times 10^{-17}$ & 2.063$\times 10^{-17}$ & 1.402$\times 10^{-16}$\\
0.200 & 1.669$\times 10^{-16}$ & 3.775$\times 10^{-17}$ & 3.575$\times 10^{-16}$ & 2.054$\times 10^{-16}$ & 1.328$\times 10^{-15}$\\
0.250 & 1.494$\times 10^{-14}$ & 2.055$\times 10^{-15}$ & 3.381$\times 10^{-14}$ & 2.043$\times 10^{-14}$ & 1.203$\times 10^{-13}$\\
0.300 & 4.493$\times 10^{-13}$ & 4.032$\times 10^{-14}$ & 1.073$\times 10^{-12}$ & 6.783$\times 10^{-13}$ & 3.737$\times 10^{-12}$\\
0.400 & 6.080$\times 10^{-11}$ & 2.643$\times 10^{-12}$ & 1.616$\times 10^{-10}$ & 1.107$\times 10^{-10}$ & 5.542$\times 10^{-10}$\\
0.500 & 1.892$\times 10^{-09}$ & 5.287$\times 10^{-11}$ & 5.603$\times 10^{-09}$ & 4.119$\times 10^{-09}$ & 1.922$\times 10^{-08}$\\
0.600 & 2.502$\times 10^{-08}$ & 7.703$\times 10^{-10}$ & 8.268$\times 10^{-08}$ & 6.471$\times 10^{-08}$ & 2.853$\times 10^{-07}$\\
0.700 & 1.901$\times 10^{-07}$ & 9.381$\times 10^{-09}$ & 7.032$\times 10^{-07}$ & 5.822$\times 10^{-07}$ & 2.447$\times 10^{-06}$\\
0.800 & 9.848$\times 10^{-07}$ & 8.302$\times 10^{-08}$ & 4.088$\times 10^{-06}$ & 3.562$\times 10^{-06}$ & 1.435$\times 10^{-05}$\\
0.900 & 3.858$\times 10^{-06}$ & 5.333$\times 10^{-07}$ & 1.803$\times 10^{-05}$ & 1.646$\times 10^{-05}$ & 6.385$\times 10^{-05}$\\
1.000 & 1.224$\times 10^{-05}$ & 2.633$\times 10^{-06}$ & 6.457$\times 10^{-05}$ & 6.152$\times 10^{-05}$ & 2.305$\times 10^{-04}$\\
1.100 & 3.297$\times 10^{-05}$ & 1.056$\times 10^{-05}$ & 1.967$\times 10^{-04}$ & 1.949$\times 10^{-04}$ & 7.069$\times 10^{-04}$\\
1.200 & 7.806$\times 10^{-05}$ & 3.593$\times 10^{-05}$ & 5.270$\times 10^{-04}$ & 5.412$\times 10^{-04}$ & 1.904$\times 10^{-03}$\\
1.300 & 1.672$\times 10^{-04}$ & 1.073$\times 10^{-04}$ & 1.273$\times 10^{-03}$ & 1.351$\times 10^{-03}$ & 4.617$\times 10^{-03}$\\
1.400 & 3.325$\times 10^{-04}$ & 2.896$\times 10^{-04}$ & 2.823$\times 10^{-03}$ & 3.088$\times 10^{-03}$ & 1.026$\times 10^{-02}$\\
1.500 & 6.329$\times 10^{-04}$ & 7.223$\times 10^{-04}$ & 5.840$\times 10^{-03}$ & 6.566$\times 10^{-03}$ & 2.121$\times 10^{-02}$\\
1.600 & 1.193$\times 10^{-03}$ & 1.696$\times 10^{-03}$ & 1.141$\times 10^{-02}$ & 1.315$\times 10^{-02}$ & 4.125$\times 10^{-02}$\\
1.800 & 4.578$\times 10^{-03}$ & 8.207$\times 10^{-03}$ & 3.826$\times 10^{-02}$ & 4.582$\times 10^{-02}$ & 1.346$\times 10^{-01}$\\
2.000 & 1.940$\times 10^{-02}$ & 3.443$\times 10^{-02}$ & 1.135$\times 10^{-01}$ & 1.391$\times 10^{-01}$ & 3.779$\times 10^{-01}$\\
2.500 & 4.658$\times 10^{-01}$ & 6.639$\times 10^{-01}$ & 1.205$\times 10^{+00}$ & 1.476$\times 10^{+00}$ & 3.201$\times 10^{+00}$\\
3.000 & 4.512$\times 10^{+00}$ & 5.784$\times 10^{+00}$ & 8.103$\times 10^{+00}$ & 9.673$\times 10^{+00}$ & 1.747$\times 10^{+01}$\\
3.500 & 2.429$\times 10^{+01}$ & 2.976$\times 10^{+01}$ & 3.714$\times 10^{+01}$ & 4.345$\times 10^{+01}$ & 6.961$\times 10^{+01}$\\
4.000 & 9.646$\times 10^{+01}$ & 1.146$\times 10^{+02}$ & 1.338$\times 10^{+02}$ & 1.536$\times 10^{+02}$ & 2.253$\times 10^{+02}$\\
4.500 & 3.308$\times 10^{+02}$ & 3.804$\times 10^{+02}$ & 4.237$\times 10^{+02}$ & 4.761$\times 10^{+02}$ & 6.457$\times 10^{+02}$\\
5.000 & 1.023$\times 10^{+03}$ & 1.140$\times 10^{+03}$ & 1.227$\times 10^{+03}$ & 1.348$\times 10^{+03}$ & 1.706$\times 10^{+03}$\\
6.000 & 6.938$\times 10^{+03}$ & 7.399$\times 10^{+03}$ & 7.675$\times 10^{+03}$ & 8.136$\times 10^{+03}$ & 9.358$\times 10^{+03}$\\
7.000 & 3.006$\times 10^{+04}$ & 3.133$\times 10^{+04}$ & 3.199$\times 10^{+04}$ & 3.326$\times 10^{+04}$ & 3.638$\times 10^{+04}$\\
8.000 & 9.187$\times 10^{+04}$ & 9.454$\times 10^{+04}$ & 9.583$\times 10^{+04}$ & 9.856$\times 10^{+04}$ & 1.050$\times 10^{+05}$\\
9.000 & 2.184$\times 10^{+05}$ & 2.230$\times 10^{+05}$ & 2.251$\times 10^{+05}$ & 2.300$\times 10^{+05}$ & 2.410$\times 10^{+05}$\\
10.000 & 4.333$\times 10^{+05}$ & 4.401$\times 10^{+05}$ & 4.431$\times 10^{+05}$ & 4.508$\times 10^{+05}$ & 4.673$\times 10^{+05}$\\
12.500 & 1.440$\times 10^{+06}$ & 1.450$\times 10^{+06}$ & 1.454$\times 10^{+06}$ & 1.470$\times 10^{+06}$ & 1.499$\times 10^{+06}$\\
15.000 & 3.093$\times 10^{+06}$ & 3.097$\times 10^{+06}$ & 3.097$\times 10^{+06}$ & 3.122$\times 10^{+06}$ & 3.154$\times 10^{+06}$\\
17.500 & 5.206$\times 10^{+06}$ & 5.191$\times 10^{+06}$ & 5.179$\times 10^{+06}$ & 5.213$\times 10^{+06}$ & 5.229$\times 10^{+06}$\\
20.000 & 7.558$\times 10^{+06}$ & 7.509$\times 10^{+06}$ & 7.479$\times 10^{+06}$ & 7.518$\times 10^{+06}$ & 7.500$\times 10^{+06}$\\
25.000 & 1.233$\times 10^{+07}$ & 1.218$\times 10^{+07}$ & 1.209$\times 10^{+07}$ & 1.213$\times 10^{+07}$ & 1.200$\times 10^{+07}$\\
30.000 & 1.664$\times 10^{+07}$ & 1.635$\times 10^{+07}$ & 1.621$\times 10^{+07}$ & 1.622$\times 10^{+07}$ & 1.594$\times 10^{+07}$\\
35.000 & 2.025$\times 10^{+07}$ & 1.981$\times 10^{+07}$ & 1.962$\times 10^{+07}$ & 1.957$\times 10^{+07}$ & 1.914$\times 10^{+07}$\\
40.000 & 2.316$\times 10^{+07}$ & 2.256$\times 10^{+07}$ & 2.235$\times 10^{+07}$ & 2.220$\times 10^{+07}$ & 2.161$\times 10^{+07}$\\
50.000 & 2.719$\times 10^{+07}$ & 2.630$\times 10^{+07}$ & 2.609$\times 10^{+07}$ & 2.566$\times 10^{+07}$ & 2.480$\times 10^{+07}$\\
60.000 & 2.945$\times 10^{+07}$ & 2.829$\times 10^{+07}$ & 2.815$\times 10^{+07}$ & 2.737$\times 10^{+07}$ & 2.630$\times 10^{+07}$\\
70.000 & 3.051$\times 10^{+07}$ & 2.913$\times 10^{+07}$ & 2.910$\times 10^{+07}$ & 2.796$\times 10^{+07}$ & 2.673$\times 10^{+07}$\\
80.000 & 3.082$\times 10^{+07}$ & 2.925$\times 10^{+07}$ & 2.934$\times 10^{+07}$ & 2.785$\times 10^{+07}$ & 2.650$\times 10^{+07}$\\
90.000 & 3.063$\times 10^{+07}$ & 2.891$\times 10^{+07}$ & 2.912$\times 10^{+07}$ & 2.732$\times 10^{+07}$ & 2.588$\times 10^{+07}$\\
100.000 & 3.014$\times 10^{+07}$ & 2.830$\times 10^{+07}$ & 2.862$\times 10^{+07}$ & 2.655$\times 10^{+07}$ & 2.505$\times 10^{+07}$\\
\noalign{\smallskip}\hline
\end{longtable}
\tablefoot{
\tablefoottext{$\dagger$}{Set IX is adopted for the results presented in Fig.~\ref{fig:mass}.}
}
}
\begin{appendix}
\label{sec:appendix}
\section{Details of $R$-Matrix calculations}
We calculated the $^{18}$F(p,\,$\alpha$) stellar reaction rate using $R$-Matrix formalism with \texttt{AZURE2} \citep{2010PhRvC..81d5805A,2014NIMPA.767..359M}.
We fixed the channel radius $a=5.25$~fm, similar to the values adopted in previous works of 5.5, 5.0, and 5.2~fm, respectively \citep{2009PhRvC..79a5801D,2011PhRvC..83d2801B,2019PhRvL.122e2701H}.
The sensitivity of the $R$-Matrix calculations on the assumed channel radius was previously explored over the range 4.5--6.5~fm (see Fig.~6 of \cite{2009PhRvC..79a5801D}), where a larger channel radius was seen to push a destructive interference dip to lower energies, and vice versa, all other parameters being equal.
We separate our employed $^{19}$Ne resonance parameters into three categories, as follows:
\begin{itemize}
\item Strongly influential resonances used in all calculations where variation of their properties within experimental uncertainties has a negligible influence on the reaction rate, shown in Table~\ref{tab:sfactor}.
\item $\ell>0$ resonances with poorly-known properties which nevertheless exhibit a minimal influence on the present uncertainty in the reaction rate, shown in Table~\ref{tab:sfactor2}.
\item $\ell=0$ resonances that dominate the present uncertainty in the reaction rate, shown in Table~\ref{tab:resdata}.
\end{itemize}

\begin{figure}
   \centering
   \includegraphics[width=9.5cm, clip=true, trim=0 0 0 30]{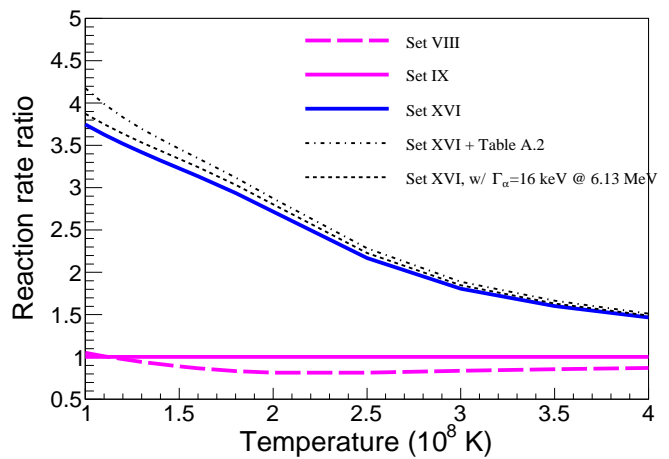}
      \caption{Comparison of the potential contribution of $\ell>0$ resonances in Table~\ref{tab:sfactor2} to the $^{18}$F(p,\,$\alpha$) rates utilized in the astrophysical calculations in the present work (normalized to our Set IX).  We demonstrate that the potential influence of the sum of those four resonances is not much more than, e.g.,  the uncertainty in $\Gamma_{\alpha}$ of the 6.13~MeV state observed by \cite{2019EPJA...55....4K}.
              }
         \label{fig:rate_nonL0}
   \end{figure}

The possible contribution of the $\ell>0$ resonances in Table~\ref{tab:sfactor2} is shown in Fig.~\ref{fig:rate_nonL0}.
We compare the sum of these four resonances against the maximum rate (Set XVI), as well as the case when the 6.13~MeV state is $1/2^{+}$ and has $\Gamma_{\alpha}=16$~keV \citep{2017PhRvC..96d4317T} rather than 8.6~keV (as assumed in Set XVI).
Our analysis showed the largest resonant contribution in Table~\ref{tab:sfactor2} is from the 6.459~MeV state in $^{19}$Ne, which was tentatively assigned as ($5/2^{-}$) by \cite{2013PhRvL.110c2502L}, who did not observe any $3/2^{+}$ states in $^{19}$Ne in this region.
As there seemed to be no corresponding $5/2^{-}$ state in $^{19}$F near this energy, \cite{2013PhRvL.110c2502L} scaled the properties of the 6.528~MeV $3/2^{+}$ state in $^{19}$F given by \cite{2007PhRvC..75e5809N}.
As our calculations already assign a mirror partner to the 6.528~MeV state in $^{19}$F, we infer that the maximum contribution of the 6.459~MeV state in $^{19}$Ne is significantly less than Fig.~\ref{fig:rate_nonL0} suggests.

For the $(3/2)^{-}$ state at 6.416~MeV, \cite{2019PhRvL.122e2701H} have the value of $\Gamma_{\rm p}$ shown in Table~\ref{tab:sfactor2}, while \cite{2020PhRvC.102d5802H} claim $\Gamma_{\rm p}=2.8\times 10^{-27}$~keV for a 1~keV resonance energy shift (inconsistent with \cite{2015PhLB..751..311B}).
We instead found $\Gamma_{\rm p}=2.3\times10^{-45}$~keV for such a $(3/2)^{-}$ resonance at $E_{\rm c.m.}=6$~keV.
We tested all of these values for $\Gamma_{\rm p}$, and even with an erroneous increase of $10^{18}$ in $\Gamma_{\rm p}$, the contribution of this state to the maximum reaction rate of Set XVI here remains negligible. 
We explore the energies near 6.416 vs. 6.423~MeV for a $(3/2^{+})$ state in Table~\ref{tab:resdata}, with appropriate shifts in $\Gamma_{\rm p}$ and consistent use of $\Gamma_{\alpha}$ so that readers can evaluate the impact of the state newly suggested by \cite{2019PhRvL.122e2701H}; this is not the same as considering a 6.416-MeV, negative parity state the same as the mirror pair of a $3/2^{+}$ state, but rather a previously unresolved doublet.

\begin{figure}
   \centering
   \includegraphics[width=9.5cm, clip=true, trim=0 0 0 30]{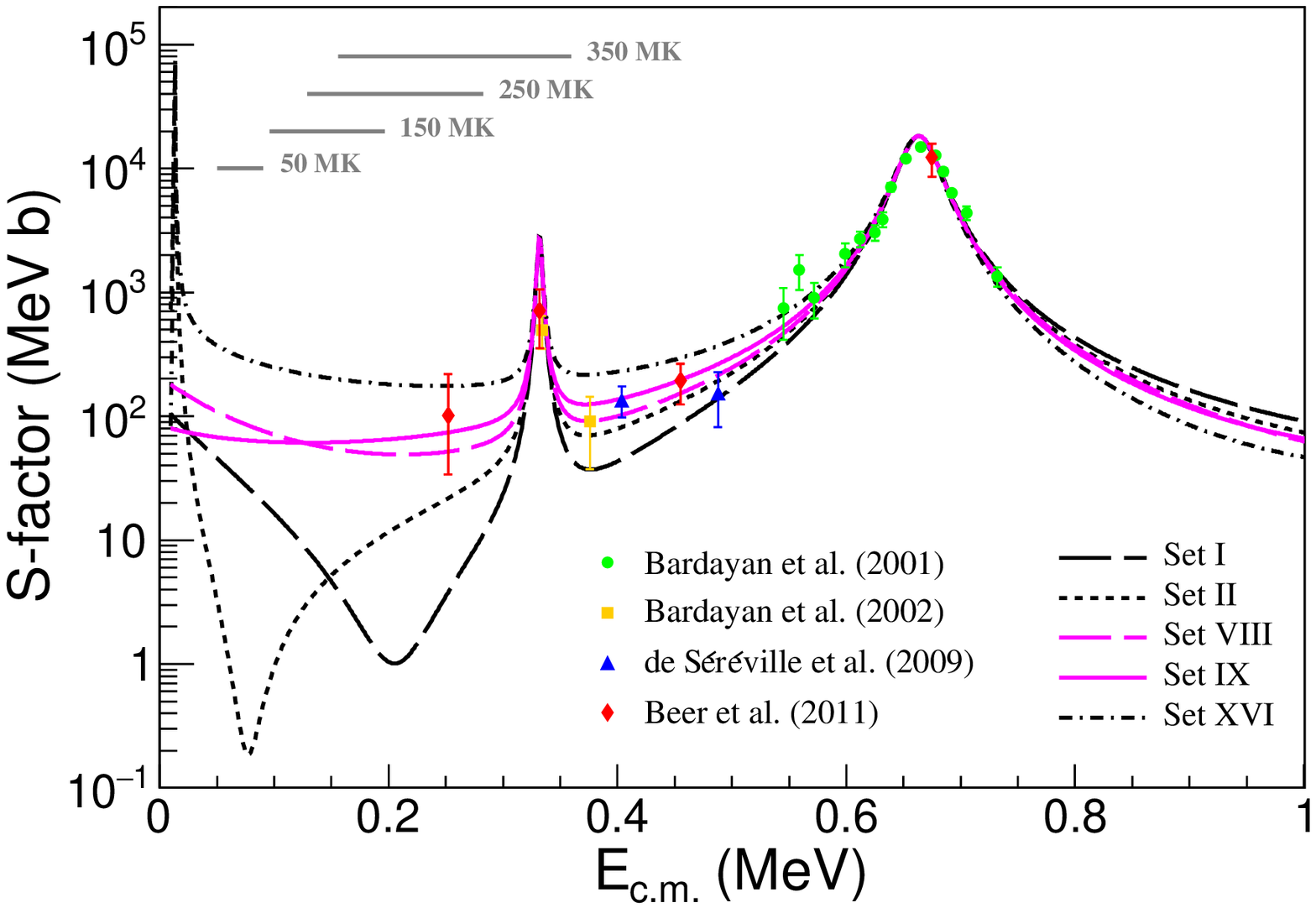}
      \caption{Astrophysical $S(E)$ for the $^{18}$F(p,\,$\alpha$) reaction.  See Tables~\ref{tab:nuclear}, \ref{tab:sfactor} and \ref{tab:resdata} for details.  Experimental works are from \cite{2001PhRvC..63f5802B,2002PhRvL..89z2501B,2009PhRvC..79a5801D,2011PhRvC..83d2801B}. 
              }
         \label{fig:sfactor}
   \end{figure}

The astrophysical S-factor, $S(E)$, is shown in Fig.~\ref{fig:sfactor} and compared with experimental results.
We show the exact $S(E)$ without taking into account the individual experimental target thicknesses, which differ from one another by more than a factor of two.
Considering the scatter in the experimental data, the agreement is reasonably good, except at 250~keV, where only two events (and no background) were observed by \cite{2011PhRvC..83d2801B}, allowing for the possibility that the true $S(E)$ could be significantly smaller as explored in the present work.

\begin{table}                                                                                                                                                                                  
 \caption{\label{tab:sfactor}
Resonance parameters used in all of the $^{18}$F($p$,\,$\alpha$) astrophysical calculations. 
}
\centering
 \begin{tabular}{lcccc}
      \hline\noalign{\smallskip}
      $E_{\rm ex}$& $E_{\rm c.m.}$  & $2J^{\pi}$ & $\Gamma_{\rm p}$     & $\Gamma_{\alpha}$  \\  
      (MeV)          & (keV)           &           & (keV) & (keV) \\  
\noalign{\smallskip}\hline\noalign{\smallskip}
      6.742  ~~\tablefootmark{a}    & 332             & $3^{-}$ & $2.22 \times 10^{-3}$   & 5.2 \\
      7.0747 ~~\tablefootmark{b}    & 664.7            & $3^{+}$ & 15.2                   & 23.8 \\
      7.79(3)~~\tablefootmark{c}    & 1380         & $1^{+}$ & 83$^{+56}_{-82}$           & 47$^{-46}_{+92}$ \\  
\noalign{\smallskip}\hline
 \end{tabular}
\tablefoot{
\tablefoottext{a}{All level parameters taken from \cite{2002PhRvL..89z2501B}} 
\tablefoottext{b}{All level parameters taken from \cite{2001PhRvC..63f5802B}} 
\tablefoottext{c}{All level parameters taken from \cite{2019EPJA...55....4K}}  
}
\end{table}

\begin{table}                                                                                                                                                                                  
 \caption{\label{tab:sfactor2}
Resonance parameters added to the maximum rate (Set XVI) exhibiting minimal influence.
}
\centering
 \begin{tabular}{ccccc}
      \hline\noalign{\smallskip}
      $E_{\rm ex}$& $E_{\rm c.m.}$  & $2J^{\pi}$ & $\Gamma_{\rm p}$     & $\Gamma_{\alpha}$  \\  
      (MeV)          & (keV)           &           & (keV) & (keV) \\  
\noalign{\smallskip}\hline\noalign{\smallskip}
6.416(4) & 6 & $(3)^{-}$ & $4.7\times10^{-50}$~~\tablefootmark{a} & $<0.5$ \\
6.439(3) & 29 & $1^{-}$ & $\leq 3.8\times10^{-19}$ & 220 \\
6.459(5) & 49 & $(5^{-})$ & $8.4\times10^{-14}$~~\tablefootmark{b} & 5.5~~\tablefootmark{b} \\
6.669(3) & 289 & $(5^{+})$ & $<2.4\times10^{-5}$ & 1.2 \\

\noalign{\smallskip}\hline
 \end{tabular}
\tablefoot{
All level parameters are taken from \cite{2019PhRvL.122e2701H}, modified here to show limit signs and tentative assignment notation provided in the original references therein. \\
\tablefoottext{a}{Differs from \cite{2020PhRvC.102d5802H}.  See the text.}
\tablefoottext{b}{\cite{2013PhRvL.110c2502L}, shifted from the $3/2^{+}$ 6.528-MeV state in $^{19}$F.  Provided for illustrative purposes only.  See the text.}
}
\begin{flushleft}
\end{flushleft}
\end{table}

\begin{table*}
\renewcommand\normalsize{\tiny}
\normalsize
\caption{\label{tab:resdata}
Resonance parameters varied in the calculation of the $^{18}$F(p,\,$\alpha$) reaction rate and $S(E)$. 
}
\centering
\begin{tabular}{ccc|cccc|cccc|cccc} \hline \hline

Set &\multicolumn{2}{c|}{Interference Signs}	& \multicolumn{4}{c|}{Properties of $\ell=0$ state}	& \multicolumn{4}{c|}{Properties of $\ell=0$ state} &\multicolumn{4}{c}{Properties of ($\ell=0$) state} \\

&$J^{\pi}$&$J^{\pi}$&$E_{\rm ex}$&$J^\pi$&ANC$_{\rm p}$&$\Gamma_\alpha$&$E_{\rm ex}$&$J^\pi$&ANC$_{\rm p}$&$\Gamma_\alpha$&$E_{\rm ex}$&$J^\pi$&$\Gamma_{\rm p}$&$\Gamma_\alpha$ \\ 
    &$1/2^+$&$3/2^+$			&(MeV)	&&(fm$^{-1/2}$)&(keV)&(MeV)	&	&(fm$^{-1/2}$)&(keV)&(MeV)	&	&(keV)		&(keV)\\
 \hline 
I	&$(++)$&$(++)$	&6.132	&$1/2^{+}$	&4	&8.6	&6.289	&$3/2^{+}$	&59	&11.6	&---	&---	&---		&---  \\
II	&$(++)$&$(+++)$	&6.132	&$1/2^{+}$	&2.5	&8.6	&6.289	&$3/2^{+}$	&59	&1	&6.423	&$3/2^{+}$	&$<3.9\times10^{-29}$	&$<0.5$ \\ 
III	&$(++)$&$(+++)$	&6.132	&$1/2^{+}$	&2.5	&8.6	&6.289	&$3/2^{+}$	&59	&1	&6.416	&$3/2^{+}$	&$<4.2\times10^{-45}$	&$<0.5$ \\ 
IV	&$(++)$&$(+++)$	&6.132	&$1/2^{+}$	&8	&8.6	&6.289	&$3/2^{+}$	&59	&11.6	&6.423	&$3/2^{+}$	&$<3.9\times10^{-29}$	&$<0.5$ \\ 
V	&$(++)$&$(+++)$	&6.132	&$1/2^{+}$	&8	&8.6	&6.289	&$3/2^{+}$	&59	&1	&6.416	&$3/2^{+}$	&$<4.2\times10^{-45}$	&$<0.5$ \\ 
VI	&$(++)$&$(++)$	&6.132	&$1/2^{+}$	&8	&8.6	&6.289	&$3/2^{+}$	&59	&1	&---	&---   		&---   		&---  \\
VII	&$(++)$&$(+++)$	&6.132	&$3/2^{+}$	&6	&0.74	&6.289	&$1/2^{+}$	&83.5	&11.7	&6.416	&$3/2^{+}$	&$<4.2\times10^{-45}$	&$<0.5$ \\ 
VIII	&$(++)$&$(++)$	&6.132	&$3/2^{+}$	&6	&0.74	&6.289	&$1/2^{+}$	&83.5	&11.7	&---	&---   		&--- 		&---  \\ 
IX	&$(-+)$&$(-+)$	&6.132	&$1/2^{+}$	&8	&8.6	&6.289	&$3/2^{+}$	&59	&1	&---	&---   		&---   		&---  \\ 
X	&$(-+)$&$(--+)$	&6.132	&$1/2^{+}$	&8	&8.6	&6.289	&$3/2^{+}$	&59	&1	&6.416	&$3/2^{+}$	&$<4.2\times10^{-45}$	&$<0.5$ \\ 
XI	&$(-+)$&$(-+)$	&6.132	&$3/2^{+}$	&6	&0.74	&6.289	&$1/2^{+}$	&83.5	&11.7	&---	&---   		&---   		&---  \\ 
XII	&$(-+)$&$(--+)$	&6.132	&$3/2^{+}$	&6	&0.74	&6.289	&$1/2^{+}$	&83.5	&11.7	&6.416	&$3/2^{+}$	&$<4.2\times10^{-45}$	&$<0.5$ \\ 
XIII	&$(-+)$&$(-+)$	&6.132	&$3/2^{+}$	&6	&0.74	&6.289	&$1/2^{+}$	&83.5	&16	&---	&---   		&---   		&---  \\ 
XIV	& ---  &$(-+)$	&---	&---    	&---	&---	&6.289	&$3/2^{+}$	&59	&11.6	&---	&--- 	   	&---  	 	&---  \\ 
XV	&$(-+)$&$(--+)$	&6.132	&$3/2^{+}$	&6	&8.5	&6.289	&$1/2^{+}$	&83.5	&16	&6.423	&$3/2^{+}$	&$<3.9\times10^{-29}$	&$<0.5$ \\ 
XVI	&$(-+)$&$(--+)$	&6.132	&$1/2^{+}$	&8	&8.6	&6.282	&$3/2^{+}$	&59	&11.6	&6.423	&$3/2^{+}$	&$<3.9\times10^{-29}$	&$<0.5$ \\ 

\hline \hline
\end{tabular}
\tablefoot{
See Table~\ref{tab:nuclear}.
}
\end{table*}

\section{Reaction rate table}
\label{sec:appendix2}

\end{appendix}

\end{document}